\pdfoutput=1

\documentclass{emulateapj}
\def\J{\hbox{J1420$-$0545$\!$}}
\def\D{\hbox{\sc DYNAGE$\!$}}

\def\G{\hbox{\sc GLEAM$\!$}}
\def\V{\hbox{VLA$\!$}}
\def\T{\hbox{TGSS ADR1$\!$}}
\def\S{\hbox{Section$\!$}}

\def\degr{\hbox{$^\circ$}}
\def\arcmin{\hbox{$^\prime$}}

\def\fs{\hbox{$.\!\!^{\rm s}$}}
\def\fdg{\hbox{$.\!\!^\circ$}}
\def\farcm{\hbox{$.\mkern-4mu^\prime$}}
\def\farcs{\hbox{$.\!\!^{\prime\prime}$}}

\shorttitle{Reassessment \J}
\shortauthors{Jamrozy et al.}

\begin{document}

\title{Reassessment of an origin of the radio structure of J1420$-$0545}

\author{M. Jamrozy$^1$, J. Machalski$^1$, B. Nikiel-Wroczy\'{n}ski$^{1}$, and H.~T. Intema$^2$}

\affil{$^1$Astronomical Observatory, Jagiellonian University, ul. Orla 171, 30-244 Krak\'ow, Poland}
\affil{$^2$Leiden Observatory, Leiden University, Niels Borhweg 2, NL-2333CA, Leiden, The Netherlands} 

\email{jamrozy@oa.uj.edu.pl, machalsk@oa.uj.edu.pl, iwan@oa.uj.edu.pl, intema@strw.leidenuniv.nl}

\begin{abstract}
In this paper, we test the possibility that the structure of the largest radio galaxy \J\,
may have been formed by restarted rather than primary jet activity. This hypothesis was motivated
by the unusual morphological properties of the source consisting of two edge-brightened, narrow,
highly collinear, and symmetric lobes, thus suggesting an almost ballistic propagation of powerful
jets into a particularly low-density external medium. New observations made with the \V\, together with 
the currently available \G\, and \T\, data releases allow the detection of an excess emission at low frequencies. 
An extracted part (88\,MHz -- 200\,MHz) of the spectrum of the emission is fitted with the \D\,
model, giving a forecast for the environmental conditions and the energetic requirements for 
the presumed old cocoon related to a preceding epoch of jet activity.
\end{abstract}


\section{Introduction}

The largest size ($\gtrsim$1 Mpc) radio galaxies are believed to have evolved from 
smaller sources; however, it is almost certain that most of the observed small sources will 
never evolve into a giant sizes. An evolutionary scheme for the general population of radio sources, 
from the smallest gigahertz peaked spectrum (GPS), through the compact steep-spectrum (CSS) ones, 
until the FRII/FRI (\citealt{Fanaroff1974}) structures, was discussed by \cite{Snellen2000}. 
Some, rather theoretical, predictions concerning this scheme was presented by \cite{Kaiser2007}. 
Most known, and well-studied, giant radio galaxies (GRGs) are 
sources of FRII-type morphology. There are a number of factors considered underlying the gigantic 
size of some radio sources: (i) high jet power, (ii) sufficiently low density of the 
ambient medium into which the jet propagates, and  (iii) the lifetime of the jet activity.  
An interesting idea is given in the paper by \cite{Subrahmanyan1996} presenting the research 
on several GRGs, where it was postulated that the giants could become very large after several 
jet activity cycles. Indeed, among the so-called double-double radio galaxies 
(DDRGs) there are a number of GRGs.

The largest known radio galaxy, \J\, (located at R.A.(J2000.0) 14$\rm^{h}$20$\rm^{m}$23$\fs$8 and 
decl.(J2000.0) $-$05$\degr$45$^{\prime}$28\farcs8), having an angular size of 17\farcm4, was discovered
by \cite{Machalski2008}. Its weak radio core coincides with a parent galaxy at a spectroscopic redshift
of $0.3067$ (presented in their paper). 
The source's double structure with a total projected linear size of $4.7$\,
Mpc is extremely slim. This structure, consisting of two opposite narrow lobes (with an
axial ratio, i.e. the ratio of the length to width of the whole radio structure, 
of about 12, measured according to a prescription given by \citealt{Leahy1984}), 
is highly collinear and symmetric; the arm ratio is only $1.08$,
and the misalignment angle is 1\fdg3. These properties are very characteristic of the 
inner lobes of known radio galaxies with a double-double structure (cf. 
\citealt{Schoenmakers2000}).

There is no doubt understanding the temporal evolution of extragalactic radio sources
involves accessing not only the jet parameters (its kinetic power, speed, lifetime) and
the properties of the ambient medium into which the lobes evolve, but also the complex
duty cycle of the jet activity. Current studies indicate that this activity can be
intermittent, or at least highly modulated on different time scales (for a review, see, 
e.g. \citealt{Saikia2009}). There is direct observational evidence that the restarted jets 
usually do not propagate through the undisturbed intergalactic medium (IGM), but instead 
within the environment substantially modified by the passage of the outflow during the 
previous stage of the jet activity (\citealt{Kaiser2000, Safouris2008}).
This may affect the observed properties of the newly formed lobes.

In a previous paper (\citealt{Machalski2011}; hereafter MJSK), the authors considered
the possibility that the apparent double structure of \J\, might be formed by restarted, rather
than primary, jet activity. The motivation for the performed analysis was the high
axial ratio of its lobes, which is the same order as -- or even higher than -- such ratios for the
inner lobes in all well-studied DDRGs, as well as the extremely low density of the external
environment, considerably young age of the structure, and the resulting very high speed of
the jet head's propagation ($\sim0.2c$, in contrast to $0.03c-0.05c$ for the inner
lobes in typical DDRGs; \citealt{Kaiser2000}; MJSK) -- where all of these properties were derived 
using the fitting procedure of the analytical model of the source's evolution.
The question was whether such a low density would be due to the unique location of \J\, in a
large void region or due to some previous jet activity epoch causing a substantial
rarefaction of the IGM. In MJSK, a model of the evolution of that
hypothetical primary structure (lobes) was assumed, and its expected luminosity and radio
spectrum were estimated. The current paper extends the observational effort to detect 
the relic's emission.

These new Karl G. Jansky Very Large Array (hereafter \V) observations of  \J\,, the relevant low-frequency 
data taken from the already available GaLactic and Extragalactic All-sky MWA Survey 
(\G\,; \citealt{Wayth2015, Hurley2017}) and the Giant Metrewave Radio Telescope (GMRT) Sky Survey 
(TGSS) First Alternative Data Release (ADR1; \citealt{Intema2017}), 
as well as the data reduction are presented in \S\, \ref{sec:data}. Fortunately, the new data allow the detection 
of an excess emission at low frequencies, which possibly originated from some old population of 
relativistic particles that may be related to the presumed cocoon (lobes), i.e. a relic of 
a previous episode of nuclear activity. The relevant data reduction and the modeling task applied 
are described in \S\, \ref{sec:mod}. A discussion of the results and final conclusions are summarized in \S\, \ref{sec:disc}.
As in previous papers, the distance, linear size, and luminosity of the analyzed source are
calculated for the $\Lambda$CDM cosmology with $\Omega_{\rm m}=0.27$, $\Omega_{\Lambda}=0.73$,
and $H_{0}=71$\,km\,s$^{-1}$\,Mpc$^{-1}$.

\section{New Radio Data}
\label{sec:data}

\subsection{VLA Observations}

The \V\, observations were carried out in the C-array configuration. The $P$-band data were
collected as a part of the project 14B--156 (PI: M. Jamrozy), consisting of three scheduling blocks
(performed in 2014 December 17, 23 and 30), each $\sim1790$ s long 
(excluding time spent for setting the antennas, and for the amplitude and phase calibrations, the total time on
the target was about 2000 s). Each of these blocks was made with the same instrumental
setup: the central frequency was set to 352 MHz and the total bandwidth was 256 MHz divided into
16 subbands, each consisting of 128 narrow channels. Such a scheme was chosen to avoid
multichannel RFI contamination. The strong and unresolved source 3C\,295 was used as the primary
calibrator, while 3C\,298 was taken as the phase calibrator. The final synthesized beam was
$63\farcs7\times 48\farcs8$ at P.A. of $-26\fdg5$.

For the data reduction, all of the \V\, datasets were imported to the Common Astronomy Software Applications package
(\textsc{casa}) and underwent a standard procedure, as outlined in the ``P-band basic data reduction''
guide.\footnote{https://casaguides.nrao.edu/index.php/0313-192\_P-band\_basic\_data\_reduction-CASA4.6} 
The most important processing steps involved ionospheric correction, flagging bad data
points (both automatic and fine-touch manual), and the calibration of flux density (according to
\citealt{Scaife2012}), delay, bandpass, gain, and instrumental polarization. All of these steps
were done on the primary calibrator, and the corrections were then transferred to the target
field. All three blocks were concatenated  and deconvolved using the \textsc{clean} algorithm.
The resulting map was used to start a self-calibration loop, but as there was no improvement --
either qualitative or quantitative -- no matter the scheme (phase only, amplitude and phase),
it was not carried on.

\subsection{GLEAM and \T\, Data}

The (\G) survey, performed with the Murchison Widefield Array 
(MWA; \citealt{Lonsdale2009}; \citealt{Tingay2013}) 
and overlying the entire sky south of decl. $+25\degr$, is described in detail by \cite{Wayth2015}. 
The survey covers the radio frequency range between 72 and 231 MHz, divided
into five bands (with the central frequencies of 87.5, 118.5, 154.5, 185.0, and 215.7 MHz), 
providing nearly continuous coverage but avoiding the band around 137 MHz, which is contaminated
by satellite interference. The angular resolution of the survey is
$2\farcm5\times 2\farcm2\sec(\delta +26\fdg7)$ at 154 MHz. The output of the first year of \G\, comprises
both the postage stamp FITS maps and the radio source catalog published by \cite{Hurley2017}.
A set of 20 images with the bandwidth of 7.7 MHz and a wideband image within 170--231 MHz
(with a resolution of $\sim 2\arcmin$) is provided. In addition, a set of three stacked maps
covering the ranges 72--103, 103--134, and 139--170 MHz is also available. All of the data (images
and catalog) are publicly accessible on the MWA Telescope website.\footnote{
http://www.mwatelescope.org/science/gleam-survey}

The \T\, survey was conducted at 150 MHz between 2010 and 2012. All of its archival raw data were
reprocessed with a fully automated pipeline based on the Source Peeling and Atmospheric
Modeling (\textsc{spam}) package (\citealt{Intema2009}; \citealt{Intema2014}), which includes a direction-dependent
calibration, modeling, and tasks correcting mainly ionospheric dispersive phase delay. The \T\, 
(\citealt{Intema2017}) includes continuum Stokes $I$ images of
the radio sky north of decl. $-53\degr$, with the resolution of $25\arcsec\times 25\arcsec$ north of
decl. $19\degr$, and $25\arcsec\times 25\arcsec\sec(\delta -19\degr)$ south of decl. $19\degr$. The noise
level is usually below 5\,mJy\,beam$^{-1}$. \cite{Intema2017} described the details of the data
processing and publicly available products (images and catalog).\footnote{http://tgssadr.strw.leidenuniv.nl/}
However, it is worth noting that the \T\, data are compiled excluding the short visibilities within a 0.2 k$\lambda$ 
distance of the $(u,v)$-plane coverage of the GMRT baselines. Its impact on the modeling procedure is explored in 
\S\, \ref{subsec:data}.

\section{Data reduction and modeling}
\label{sec:mod}

\subsection{New Radio Maps and the Spectrum of \J}
\label{subsec:data}

\begin{figure*}
\centering
\includegraphics[angle=0, scale=0.65]{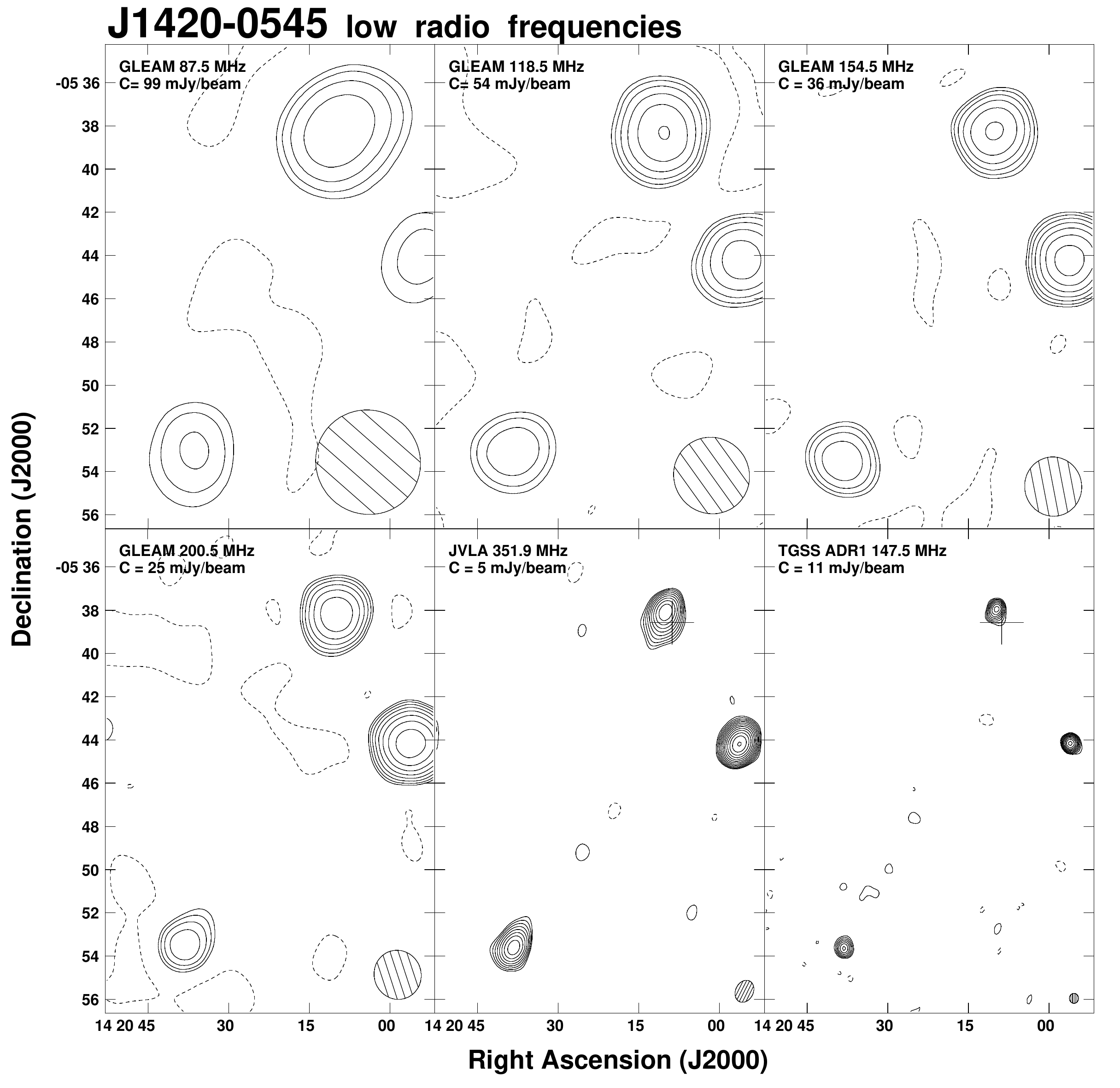}
\caption{Radio maps of \J\, (the northern and southern lobes) observed at six different
frequencies: 87.5 MHz, 118.5 MHz, 154.5 MHz, and 200.5 MHz from \G\,, as well as at 147.5 MHz
from \T\, and at 351.9 MHz from \V\,. The contours, spaced in factors of $\sqrt{2}$ in brightness, 
are plotted starting with the value C ($\simeq 3 \times$ the rms noise in the relevant map; cf. 
Table\,\ref{tab:observations}) given in each panel in units of mJy\,beam$^{-1}$. 
The cross marks the position of the background confusing source. The sizes of the beams 
are indicated by the hatched circles/ellipses in the bottom-right corner of each image.}
\label{fig:all}
\end{figure*}

The new low-frequency maps of \J\, obtained from the above data are compiled in Fig.\,\ref{fig:all}, 
and the instrumental characteristics of all radio maps included in this paper are collected in 
Table\,\ref{tab:observations}.

Further steps in the data
reduction consisted of the (i) correction of the flux densities of the narrow opposite lobes for the
emission of the compact background source that is confusing the northern lobe and is clearly visible on the
high-resolution maps in MJSK (their Fig.\,2), (ii) determination of the
corrected radio spectrum of these lobes, and (iii) separation of their low-frequency emission from 
that detected in the \G\, survey. In the first step, the spectrum of the confusing source 
given in MJSK (column\,7 in their Table\,4) was extrapolated to the frequency of 148 MHz, with the slope of 0.58
arising from its flux densities at 329 and 619 MHz and used
to correct the flux densities measured at the \V\, and 
\T\, maps where its contribution is not resolved (the confusing source position on the \V\, and \T\,
maps is marked in Figure \ref{fig:all}). In order to determine the spectrum of the slim
lobes' structure at frequencies below $\sim$150 MHz, a synchrotron model was fitted to those corrected
flux densities but excluding the \G\, data. This is done in order to separate the excess emission in the
low-resolution \G\, frequencies relative to the compact emission detected with the high-resolution observations 
at frequencies above $\sim$150 MHz. Because of the significant steepness of this spectrum at frequencies above
1400 MHz, the JP (\citealt{Jaffe1973}) model was used to evaluate the flux densities of these lobes at the \G\,
frequencies. 
The fitted models, including or excluding the \T\, data point, appear almost identical -- the flux 
densities expected at the \G\, frequencies differ by less than 1 mJy.
Finally, the flux densities determined this way were subtracted from
those resulting from the \G\, survey. The latter ones are the flux densities of the entire radio structure within 
the four frequency bands [72--103], [103--134], [139--170], and [170--231] MHz, and were derived by averaging their 
values provided in the \cite{Hurley2017} catalog with our Gaussian-model fits performed on the FITS maps with the 
Astronomical Image Processing System \textsc{aips} task \textsc{jmfit}.

\begin{table}[]
\footnotesize
\centering
\caption{Instrumental characteristics of the presented radio maps}
\begin{tabular}{lccrc}
\hline
\hline
Telescope/     & Frequency    & Beam Size      & P.A.   & rms Noise \\
Survey         & $\nu$ (MHz)  & (arcsec)       & (deg)  & (mJy beam$^{-1}$) \\
\hline
MWA/           &  87.5        & 294$\times$287 & 48.7   & 32.9 \\
GLEAM          & 118.5        & 215$\times$209 & 35.6   & 17.8 \\
               & 154.5        & 165$\times$158 & 10.9   & 12.3 \\
               & 200.5        & 137$\times$130 & 18.1   &  8.5 \\
GMRT/          & \\
TGSS\,ADR1     & 147.5      & 27.6$\times$25.0 &  0.0   &  3.8 \\
               & \\
VLA            & 351.9      & 63.7$\times$48.8 & $-$26.5 & 1.7 \\
\hline
\end{tabular}
\label{tab:observations}
\end{table}

\begin{figure}
\centering
\includegraphics[angle=0, scale=0.65]{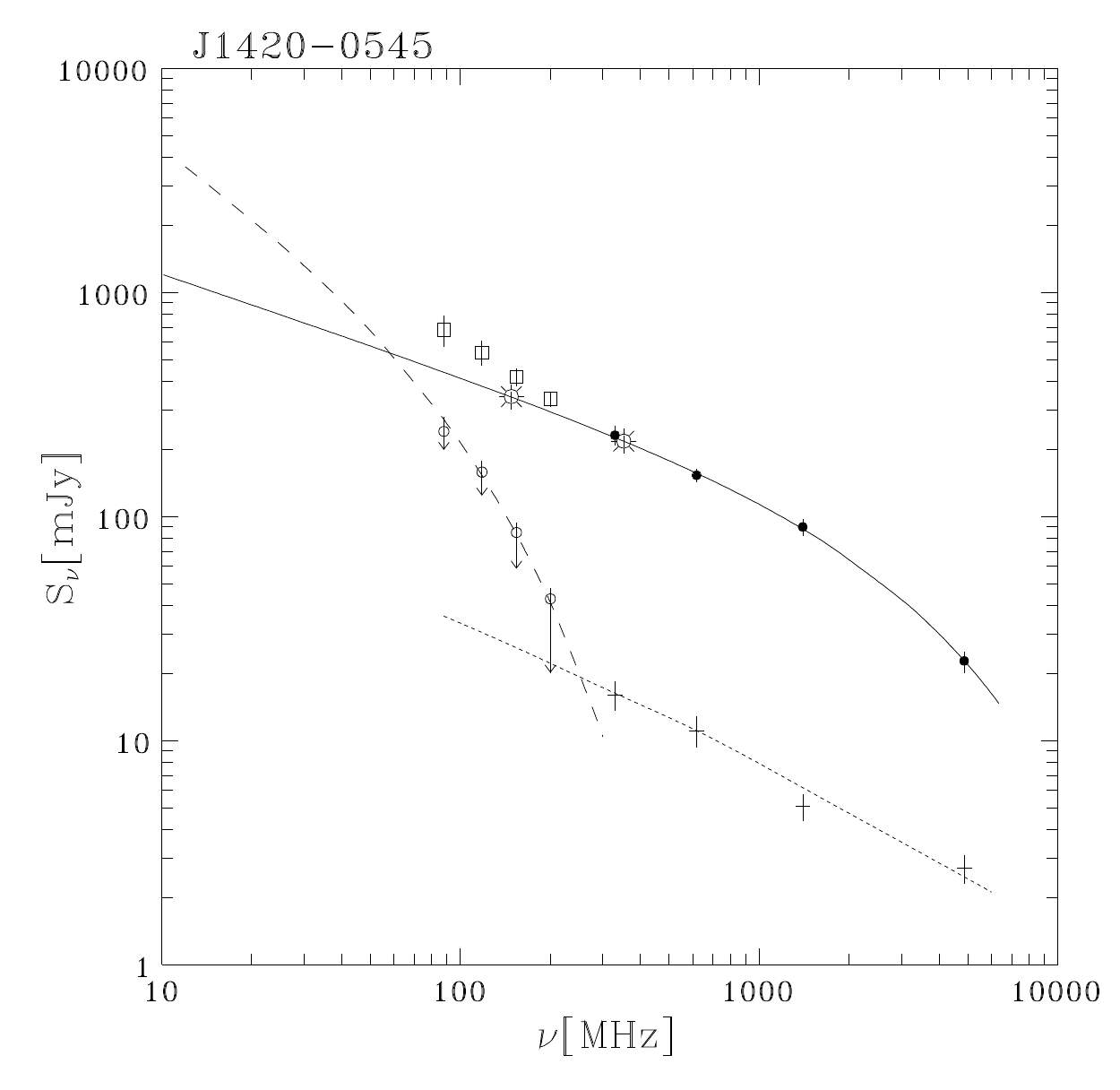}
\caption{Radio spectra of the apparent narrow lobes of \J\, fitted with the JP model (the solid line), the
faint source confusing the northern lobe (the dotted line), and the \D\, model fit to the residual data points 
originated within the low-frequency range of the spectrum from the \G\, survey (the dashed line). 
Different symbols mark the flux-density data points collected in Table\,\ref{tab:flux}: open squares -- average \G\, data (column 2), 
crosses -- confusing source (column 4), large crossed open circles and small dots -- corrected \T\,/\V\, and original MJSK data,
respectively (column 5), and small unfilled circles -- residual data (column 6). Explanations for the arrows connected to them are given 
in \S\, \ref{sec:disc}.}
\label{fig:spect}
\end{figure}

All of the above radio data are collected in Table\,\ref{tab:flux}. Columns\,2 and 3 give the above averaged \G\, 
flux densities, and those derived from the \T\, and \V\, maps, respectively. A similar
procedure is performed for the error analysis. Column\,4 gives the flux densities of the confusing source given in
MJSK and supplemented with their estimates at the frequencies of \G\, and \T\,, while column\,5 gives the flux densities
of the narrow opposite lobes dominating the high-frequency maps in MJSK. 
The flux densities in column\,6 are related to the residual emission at low frequencies, i.e., they
show the difference between the entries in columns 2 and 5, while column\,7 indicates the \D\, model fit to
these values described in \S\, \ref{sec:mod}. The spectra resulting from the above data are presented in Figure\,\ref{fig:spect}.

\begin{table*}[]
\footnotesize
\centering
\caption{New observed flux densities of the entire structure of \J\, (Columns 2 and 3) 
and the archival data for its slim lobes and the background source confusing the northern lobe 
(Columns 4 and 5 with the relevant Notes appended under the Table)}
\vspace{2mm}
\begin{tabular*}{129mm} {@{}c @{}c @{}c  @{}c  @{}c @{}c @{}c @{}c @{}c @{}c @{}c  @{}c @{}c}
\tableline
\tableline
\multicolumn{2}{c}{} &
\multicolumn{4}{c}{Entire Structure} &
\multicolumn{2}{c}{N-conf.} &
\multicolumn{2}{c}{Slim Lobes}	& 
\multicolumn{2}{c}{Residual} &
\multicolumn{1}{c}{Model} \\	
\multicolumn{1}{c}{}	&
\multicolumn{1}{c}{} &
\multicolumn{1}{c}{GLEAM}  &
\multicolumn{1}{c}{} &
\multicolumn{1}{c}{TGSS/VLA} &
\multicolumn{1}{c}{} &
\multicolumn{1}{c}{Source} &
\multicolumn{1}{c}{} &
\multicolumn{1}{c}{Corrected} &
\multicolumn{1}{c}{} &
\multicolumn{1}{c}{Flux} &
\multicolumn{1}{c}{} &
\multicolumn{1}{c}{Fit}\\
\multicolumn{1}{c}{$\nu$[MHz]}	&
\multicolumn{1}{c}{} &
\multicolumn{1}{c}{$S_{\nu}$(mJy)}  &
\multicolumn{1}{c}{} &
\multicolumn{1}{c}{$S_{\nu}$(mJy)} &
\multicolumn{1}{c}{} &
\multicolumn{1}{c}{$S_{\nu}$(mJy)} &
\multicolumn{1}{c}{} &
\multicolumn{1}{c}{$S_{\nu}$(mJy)} &
\multicolumn{1}{c}{} &
\multicolumn{1}{c}{$S_{\nu}$(mJy)} &
\multicolumn{1}{c}{} &
\multicolumn{1}{c}{$S_{\nu}$(mJy)}\\
(1) &  &  (2)  &  &  (3) &  & (4)  &  & (5) &  & (6)  &  &  (7) \\
\tableline
$\;\;\;$87.7 &  & 681$\pm$110 & &   ...        & & ...                 & & 441$\pm13\rm^{c}$  & & 240$\pm$40 & & 264.9\\
$\;\:$118.4  &  & 540$\pm$68  & &   ...        & & ...                 & & 382$\pm5\rm^{c}$   & & 158$\pm$20 & & 151.6\\
$\;\:$147.5  &  & ...         & &   369$\pm$19 & & 26$\rm^{b}$         & & 343$\pm$17         & & ...        & & ...  \\
$\;\:$154.3  &  & 420$\pm$40  & &   ...        & & ...                 & & 335$\pm4\rm^{c}$   & &  85$\pm$9  & & 84.3 \\
$\;\:$200.5  &  & 335$\pm$26  & &   ...        & & ...                 & & 293$\pm4\rm^{c}$   & &  42$\pm$5  & & 43.0 \\
$\;\:$328.8  &  & ...         & &   ...        & & 16.0$\pm2.4\rm^{a}$ & & 231$\pm24\rm^{a}$  & & ...        & & ...  \\
$\;\:$351.9  &  & ...         & &   232$\pm$13 & & 15.4$\rm^{b}$       & & 217$\pm$11         & & ...        & & ...  \\
$\;\:$617.3  &  & ...         & &   ...        & & 11.1$\pm1.8\rm^{a}$ & & 153$\pm11\rm^{a}$  & & ...        & & ...  \\
1400.0       &  & ...         & &   ...        & &  5.1$\pm0.7\rm^{a}$ & & 90$\pm8\rm^{a}$    & & ...        & & ...  \\
4860.1       &  & ...         & &   ...        & &  2.7$\pm0.4\rm^{a}$ & & 22.7$\pm2.3\rm^{a}$& & ...        & & ...  \\ 
\tableline
\end{tabular*}
\begin{flushleft}
{\bf Notes.}  Columns 6 and 7 give the low-frequency residual flux densities interpreted as the presumed relic 
emission of a former cocoon and the \D\, fit to this relic, respectively.\\
$\rm^{a}$ Flux densities from MJSK.\\ 
$\rm^{b}$ Extrapolated flux densities. \\
$\rm^{c}$ Flux densities from the JP model fit shown in Figure\,\ref{fig:spect}.
\end{flushleft}
\label{tab:flux}
\end{table*}

\begin{figure}
\centering
\includegraphics[angle=0, scale=0.65]{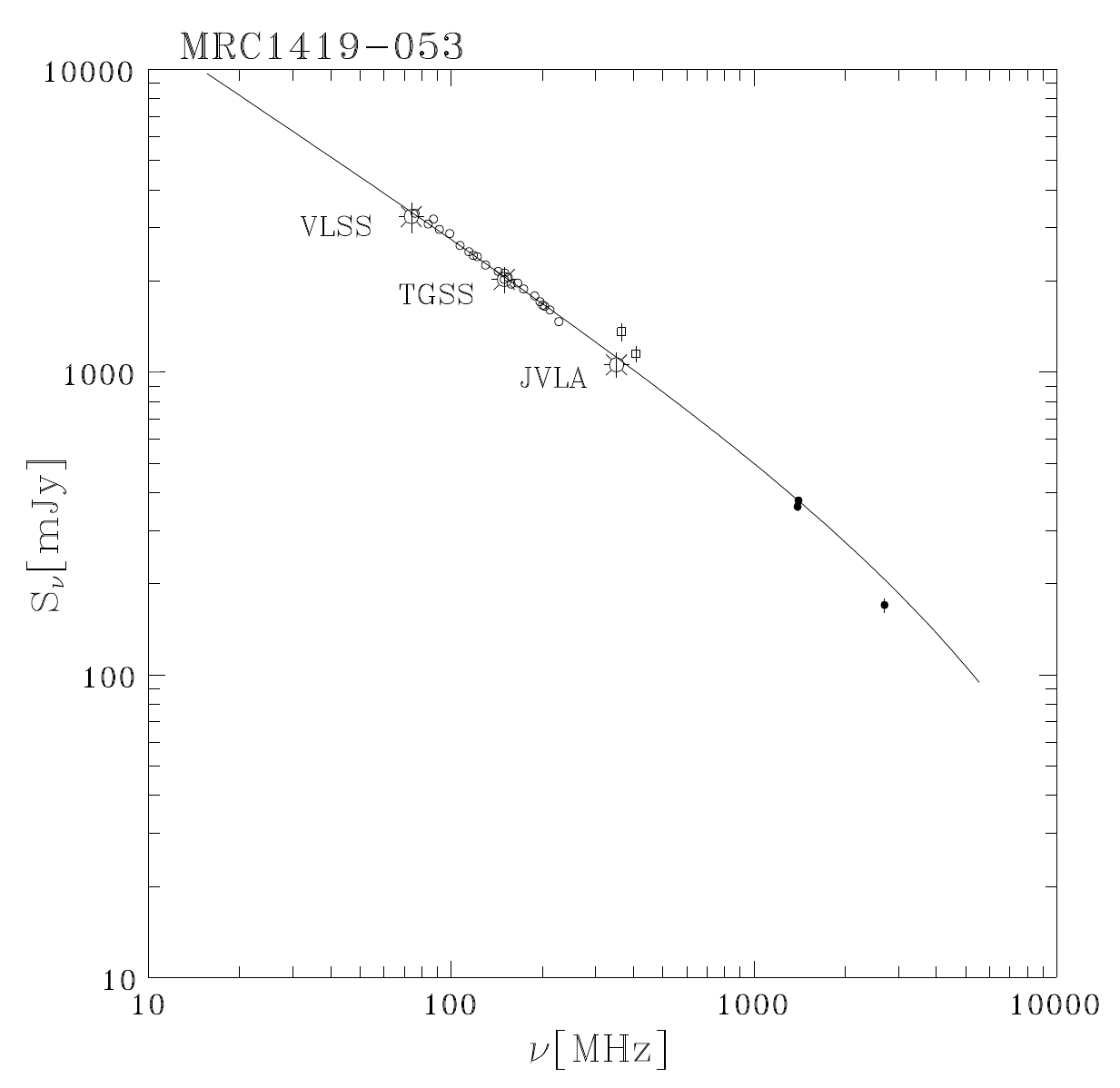}
\caption{Radio spectrum of the neighboring source MRC\,1419$-$053 used to control
the consistency of the flux-density scales at low frequencies. All of the \G\, channel data are marked by
small unfilled circles. The VLSSr, \T\,, and \V\, flux densities are shown by large crossed unfilled circles, 
while small dots indicate the FIRST (\citealt{Becker1995}), NVSS (\citealt{Condon1998}), and PKS (\citealt{Wright1990}) fluxes. 
The TXS and MRC data are marked by small unfilled squares. The solid line shows the JP model fitted to the data points.}
\label{fig:mrc}
\end{figure}

The ``residual'' flux densities, i.e. the difference between their values detected in the \G\, survey and
the flux densities estimated for the apparent lobes of \J\, are significant, suggesting an excess of the
low-frequency emission with a very steep spectral index ranging from $\alpha=1.9\pm0.03$ at 88 MHz to 
$\alpha=2.7\pm0.05$ at 200 MHz.
Such a spectrum resembles well-known spectra that are a superposition of two components: steep-spectrum extended
lobes and a flat-spectrum compact radio core, i.e. where emission of a bright core strongly flattens their
high-frequency slope. In the case of \J\,, such a bend-up appears at low frequencies. In order to check
whether this might be caused by the incompatible flux-density scales applied in the sky surveys involved,
we check the spectrum of the neighboring compact source MRC\,1419$-$053 (PKS\,B1419$-$053).
Its spectrum, shown in Figure \ref{fig:mrc}, includes the flux-density points collected from all the surveys involved,
i.e. \G\,, \T\,, and \V\,, as well as points from other databases. All data points that originated from the
above surveys and supplemented with the 74 MHz flux density from the VLA Low-Frequency Sky Survey Redux 
(VLSSr) catalog (\citealt{Lane2014}) indicate a consistency of their calibration scales. In this context, 
we note that the flux densities of the latter source are 
overestimated in the 365 MHz Texas survey (TXS; \citealt{Douglas1996}) and in the 408 MHz Molonglo Reference
Catalog (MRC; \citealt{Large1981}). Thus, we cannot exclude the possibility that the observed 
low-frequency emission excess may be related to the hypothetical cocoon that formed during a previous episode of the jet
activity precluding the one actually observed, i.e. the scenario supposed and considered in MJSK.
In that paper, it was completely hypothetical. However, the new observations presented here change the
situation; the emission excess at the low frequencies cannot be ignored. Therefore, we undertake
a further attempt to fit a dynamical model to these residual flux-density points.

\subsection{Modeling Procedure}

Similarly to the previous works (e.g. \citealp{Machalski2008,Machalski2011,Machalski2016}), the modeling procedure performed
in this paper uses the \D\, algorithm of \cite{Machalski2007}. This numerical code allows the determination, 
for a given set of observables, of the four main parameters of the dynamical model, the jet power $Q_{\rm j}$,
the density of the external gaseous medium near the radio core $\rho_{0}$, the age of the lobes' structure
$t$, and the initial exponent of the electron energy distribution injected by the jets into the lobes
$p=1+2\alpha_{\rm inj}$, where $\alpha_{\rm inj}$ is the injection spectral index, which characterizes 
the slope of the spectrum at a very young age of the radio source. However, the apparent slope of the four
residual flux-density points in Figure\,\ref{fig:spect} strongly suggests an emission of some old population of relativistic
electrons, e.g. a relic emission typical for ``dying'' (old) sources, i.e. whose activity of their nuclei was
terminated or fell down to such a low level that a jet outflow would be inefficient or disappear completely
(cf. \citealt{Komissarov1994}; \citealt{Murgia2011}).
For this reason, an extension of the \D\, code (the dynamical model with terminated jet activity, KDA EXT),
described by \cite{Kuligowska2017}, was applied. The extension of the \D\, code is based on the division of 
the cocoon's temporal evolution into two periods: the time of the jet activity until its termination
($t_{\rm j}$) and that from the terminated activity until the actual age of the source,
$t$.
The evolution during the first period is described in detail in \cite{Machalski2007}. 
A further evolution after $t_{\rm j}$, proposed by \cite{Kuligowska2017},
relying on modified expressions for the adiabatic expansion of the cocoon, its
pressure, and integrated radio power, is summarized in the Appendix.

In the first step of the modeling, we repeated the calculations of MJSK using the new data for the narrow
lobes (the entries in column 5 of Table\,\ref{tab:flux}), and realized that the earlier models for the ``Primary'' or ``Inner''
origin of the structure are not changed appreciably. In the second step, we fit the KDA EXT model to the
entries in column 6 of Table\,\ref{tab:flux}. Because of the lack of any data at frequencies below 329 MHz in MJSK, the
authors have had to assume values for two of the four model parameters predictable by the model fit, i.e. the
values of $Q_{\rm j}$ and $\rho_{0}$. In this paper, we only assume the $Q_{\rm j}$ value, equating it to
the value fitted in the ``Inner'' case solution. The other required assumptions, kept identical as in MJSK,
are summarized in Table\,\ref{tab:mod}. The last eight lines in Table\,\ref{tab:mod} show the fit results. Note that the best-fit value
of $Q_{\rm j}$ is 1.14 times higher than the initially assumed value. Such an increase was necessary to obtain a
better normalization of the modeled spectrum without changing its shape. Likewise, the central density in the
present solution is comparable to that in MJSK. Further aspects of the fit are discussed in the next section.

\begin{table}[]
\centering
\tiny
\caption{Model input parameters for the presumed old cocoon}
\vspace{2mm}
\begin{tabular}{lrr}
\tableline
\tableline
Parameter   &  Symbol   & Value \\
$\;\;\;\;\;$     (1)   &    (2)     & (3) \\
\tableline
Set: \\
Linear size             & $D$[kpc]  & 4690 \\
Axial ratio             & $R{\rm_T}$      & 6.0 \\
Jet power        & $Q_{\rm j}$[10$^{45}$erg\,s$^{-1}$] & 3.9 \\
Adiabatic index of the cocoon material  & $\Gamma_{\rm c}$  & 5/3 \\
Adiabatic index of the ambient medium   & $\Gamma_{\rm x}$  & 5/3 \\
Adiabatic index of the magnetic field   & $\Gamma_{\rm B}$  & 4/3 \\
Minimum electron Lorentz factor (injected) & $\gamma_{\rm min}$ & 1 \\
Maximum electron Lorentz factor (injected)  & $\gamma_{\rm max}$ & 10$^{7}$ \\
Core radius of the ambient density distribution        & $a_{0}$[kpc] & 10 \\
Slope of the ambient density distribution  & $\beta$ & 3/2 \\
Thermal particles within the cocoon        & $k^{\prime}$ & 10 \\
\\

\noindent
Observed:  \\
Luminosity spectral densities at a number \\
of observing frequencies, $i=1,\,2,\,3,\,4$ &  $P_{\nu_{i}}$ & {\rm Note}\\
\\
\noindent
Fit: \\
Jet power (!)     & $Q_{\rm j}$[10$^{45}$erg\,s$^{-1}$] & 4.5 \\
Central density  & $\rho_{0}$[10$^{-26}$g\,cm$^{-3}$]  & 3.9 \\
Injection spectral index  & $\alpha_{\rm inj}$  & 0.65 \\
Cocoon age       & $t$[Myr]  & 315 \\
Jet activity duration & $t_{\rm j}$[Myr] & 110 \\
External density      & $\rho_{\rm D/2}$[10$^{-29}$g\,cm$^{-3}$] & 1.1 \\
Cocoon pressure       & $p_{\rm c}$[10$^{-13}$dyn\,cm$^{-2}$]    & 3.0 \\
Goodness of the fit   & $\chi^{2}\rm_{red}$  &  0.54 \\
\tableline
\end{tabular}
\begin{flushleft}
{\bf Note.} Relevant luminosities are calculated using the flux densities given in column 6 of Table\,\ref{tab:flux},
(!) -- see the text.
\end{flushleft}
\label{tab:mod}
\end{table}

\section{Discussion and Conclusions}
\label{sec:disc}

Since the GLEAM's restoring beam sizes are much larger than the
corresponding beams in the high-frequency maps in both in MJSK and in this paper,
the residual flux densities (in column 6 of Table \ref{tab:flux}) do not account for an
additional flux which the \G\, observations might detect from eventual
steep-spectrum sources not seen in the higher-frequency bands. An upper limit on
this additional flux would be determined by fitting a JP model to the sum of entries
marked with the diamond (columns 4 and 5) and the entries in column 3. As the
result, we realize that the residual flux densities (column 6) might decrease
to $\sim$85\% at 88 MHz until $\sim$50\% at 200 MHz, as
shown by the arrows in Figure \ref{fig:spect}. 
In such a situation, the radio spectrum of the presumed old cocoon would steepen
even more than that shown by the dashed line in Figure \ref{fig:spect}. Keeping the above in
mind, we might expect that a relevant \D\, model will predict a little lower
jet power and a higher age than their values given in Table \ref{tab:mod}. In this context,
it is worth emphasizing that the radiative ages ever determined for the relic
radio sources have never exceeded 300 Myr (cf. \citealt{Slee2001}; \citealt{Murgia2011}).
Therefore, this is very likely that a low surface brightness of the old cocoon
may be hard to detect, and its brightest regions are only visible with the
\G\,.

A significant difference between the presented model of the presumed old cocoon and the ``Outer''
solution in MJSK (the last column in their Table\,6) is the high $\alpha_{\rm inj}$ value of 0.65. Such
values usually characterize the steep or very steep spectra of sources observed at high redshifts.
This is expected due to the evidence for the kinetic temperature and the density of the IGM evolution with
redshift. Therefore, at higher redshifts, the radio jet works against the higher pressure of the
denser ambient medium (cf. \citealt{Athreya1998}). In such circumstances, the propagation speed of the jet's head, 
as well as the bulk flow behind this head, would slow down. Consequently, the acceleration process would lead to
a steeper value of $p$ in the energy distribution. However, the very large size and uncommon propagation
speed of \J\, (resulting from its relatively young age determined from the model) testify against 
a dense environment. The analysis of Kuligowska (2017) shows how much the radio spectrum will steepen in relation
to its initial slope when a period of quiescence is comparable to the lifetime of the jet activity, even
if $\alpha_{\rm inj}$ is not very high. 

Another difference between the presented model and the ``Outer'' model in MJSK is
the fitted age of the old cocoon, 315 Myr vs. 178 Myr. However, both solutions were 
based on the arbitrary assumptions about the jet power during the primary nuclear
activity, $Q_{\rm j,out}\simeq Q_{\rm j,inn}$ and $Q_{\rm j,out}\simeq3\times Q_{\rm j,inn}$,
respectively. Such assumptions were necessary considering a very short part of the
radio spectrum that is suspected of being a relic, as well as simply its lack in MJSK.
Therefore, one can expect that several quite different models may fit the data points
in column\,6 of Table\,\ref{tab:flux}). 
However, the difference between the ages of the presumed old cocoon
derived within those solutions, which are comparable to the fitted time lags
$t-t_{\rm j}$ ($\sim$110 Myr and $\sim$106 Myr, respectively), suggests an
uncertainty of the ages not larger than 50\% of their fitted values.
Further observations at frequencies much lower than 70 MHz would help to constrain the variety of such models.

On the other hand, the models presented here and in MJSK provide similar very low
densities of the gas surrounding the hypothetical primary lobes of J1420$-$0545
($\rho_{\rm out}\sim 10^{-29}$ g\,cm$^{-3}$). This result, together with the density of the
new cocoon related to the observed slim lobes ($\rho_{\rm inn}\sim 2\times 10^{-31}$
g\,cm$^{-3}$) predicted in MJSK and the fact that the latter value is consistent with
the mean density of the baryonic matter in the universe, implies that the extensive
discussion of these issues given in MJSK (in their Sect.\,4.4) still remains valid.

Some doubts may arise concerning the long time lag between subsequent jet activities
implied by the model derived in this paper, and uncertainty whether such a low-density environment caused
by a primary phase of activity would remain in place for the current phase, may arise. This 
was already considered and extensively discussed by \cite{Kaiser2000}. They argued
that a replacement of the old cocoon material by the surrounding external medium
(due to the buoyancy effect and/or a possible pressure gradient within the cocoon)
proceeds at the sound speed in this medium. Therefore, although these effects undoubtedly
depend on its kinetic temperature, the entrainment of gaseous material is a slow process,
and \cite{Kaiser2000}, p. 389 concluded: ``... it seems unlikely that the environments of the inner
sources of the DDRGs are created by the replacement of the old cocoon material by the
denser IGM''. In the case of J1420$-$0545, proper X-ray observations and eventual detection 
of radial surface brightness and gas density profiles in the vicinity of its host galaxy would 
be crucial for discriminating between the two possible origins of the observed radio structure: 
(i) repeated periods of jet activity, or (ii) a unique (primary) episode of the jet 
activity appearing in a deep void of the IGM at the outskirts of a filamentary Warm-Hot IGM (WHIM).
For a comprehensive review concerning cosmic voids, see, e.g. \cite{Weygaert2011}.

To sum up, in this paper, we test the possibility that the structure of the largest radio galaxy, \J\,, discovered by 
\cite{Machalski2008}, may be formed by restarted, rather than primary, jet activity. As mentioned in
\S\, 1, this hypothesis was motivated by the unusual morphological properties of the source, suggesting an 
almost ballistic propagation of powerful jets in a particularly low-density ambient medium. 
Numerical simulations of the development of jets in a pre-existing cocoon of synchrotron emission
have been presented by \cite{Clarke1991}. They suggest that the supersonic propagation
of a restarted jet in the old cocoon can excite a weak bow shock immediately ahead of this jet. However,
this scenario seems to be inconsistent with observations of the Mpc-scale DDRGs, which show the inner double
structures as edge-brightened lobes rather than near-ballistic jets. In this context, \cite{Clarke1997}
considered the possibility that a very large Mach number may force the bow shock to hug the jet along its
length, so that the emission from both the bow shock and the restarted jet would together form a narrow
inner structure. The obtained results and the final conclusions are as follows.

\begin{enumerate}

\item
{New observations of \J\, conducted with the \V\, and the data provided from the \T\, extend the
low-frequency spectrum of its two narrow lobes. Repeated calculations in MJSK using these
supplementary data show that the earlier dynamical models for these lobes are not changed appreciably.
Different model fits performed for the observed radio structure imply a relatively young age of the source
$\sim 35$\,Myr, its relatively high expansion velocity $\sim 0.2\,c$, and large kinetic power $\sim 4 \times 
10^{45}$\,erg\,s$^{-1}$, and confirm a particularly low-density environment $\lesssim 10^{-29}$\,g\,cm$^{-3}$.}

\item
{The emission detected with the \G\, survey seems to support the hypothesis about the presence of an old cocoon
expected in the framework of the jet intermittency scenario.
It is intriguing that the brightest regions in the \G\, maps overlap the narrow lobes presumed to be formed
by a restarted jet activity. Their emission is likely contaminated by an old population of relativistic electrons 
from a previous episode of nuclear activity. The lack of detectable emission in between these regions
means that, even if exists, its radio surface brightness is too low to be discerned from the cosmic background.} 

\item
{In spite of the above, we can conclude that the dilemma of whether the extremely low density of the \J\, environment
is due to a previous jet activity or rather due to its unique location in a large void region of the galaxy
distribution is still not settled. 
Perhaps, future Low Frequency Array (LOFAR) observations in the low-frequency band (15\,MHz -- 70\,MHz) may
help in investigating the presence of a hypothetical outer cocoon in this source; however, the southern
declination of the target might cause a problem.}
\end{enumerate}

\acknowledgements
This scientific work makes use of the Murchison Radio-astronomy Observatory, 
operated by CSIRO. We acknowledge the Wajarri Yamatji people as the traditional owners of the observatory site. Support for the operation of the 
MWA is provided by the Australian Government (NCRIS), under a contract to Curtin University administered by Astronomy Australia Limited. 
We acknowledge the Pawsey Supercomputing Centre, which is supported by the Western Australian and Australian Governments.
The authors acknowledge the National Radio Astronomy Observatory, which is a facility of the National Science Foundation operated under 
cooperative agreement by Associated Universities, Inc., for observing time, as well as the staff of the GMRT that made these observations possible. 
GMRT is run by the National Centre for Radio Astrophysics of the Tata Institute of Fundamental Research.
We thank Dr. Thomas Franzen for explanations concerning the GLEAM data and the anonymous 
referee for useful suggestions that helped to improve the paper. J.M. and M.J. were supported in part by 
the Polish National Science Centre (NSC) grant DEC-2013/09/B/ST9/00599.

\appendix
%\section{Appendix}
The density distribution of the ambient gaseous medium, $\rho_{\rm a}$, (identical
for both periods: the jet activity and after its termination), is
$\rho_{\rm a}(r)=\rho_{0}(r/a_{0})^{-\beta}$, where $\rho_{0}$ is the central density
of the radio core with radius $a_{0}$, and $\beta$ is the exponent of the density profile.

\noindent
The total length of the jet $r_{\rm j}$ arising from the energy conservation conditions,
i.e. approximately one-half of the source's linear size at the time of termination,
$D(t_{\rm j})$, is

\begin{equation}
r_{\rm j}(t_{\rm j})=c_{1}\left(\frac{Q_{\rm j}}{\rho_{0}a_{0}^{\beta}}\right)
^{1/(5-\beta)}t_{\rm j}^{3/(5-\beta)}\approx\frac{D(t_{\rm j})}{2}
\end{equation}

\noindent
where $Q_{\rm j}$ is the jet's power and $c_{1}$ is a constant dependent on the
ratio of the jet head's pressure and the uniform cocoon pressure, ${\cal P}_{\rm hc}$,
as a function of the cocoon's axial ratio $R_{\rm T}$. Note that an explanation of all
parameters describing a source and its analytical model is given in Table\,\ref{tab:mod}.

The source's length at its actual age, $t\geq t_{\rm j}$, is given by

\begin{equation}
D(t,t_{\rm j})=D(t_{\rm j})\left(\frac{t}{t_{\rm j}}\right)^{c_{4}},
\end{equation} 

\noindent
where $c_{4}=\frac{2(\Gamma_{\rm c}+1)}{\Gamma_{\rm c}(7+3\Gamma_{\rm c}-2\beta)}$.

The cocoon's pressure after switches off the jet's activity is

\begin{equation}
p_{\rm c}(t>t_{\rm j})=p_{\rm c}(t_{\rm j})\left(\frac{t}{t_{\rm j}}\right)
^{-3\Gamma_{\rm c}c_{4}},
\end{equation}

\noindent
where the uniform cocoon pressure during the jet's activity, $p_{\rm c}(t_{\rm j})$,
is the same as that given by \cite{Kaiser1997a} [their Equation 20].

The analytical formula for the total ratio power, $P_{\nu}$, of a source (its cocoon) at a given
frequency is written as the sum of two integrals. The first integral gives the power
calculated until the time $t_{\rm j}$, while the second one adds the power of the emission
from $t_{\rm j}$ until the actual age of the source, $t$,

\begin{equation}
P_{\nu}(t)=\left\{ \begin{array}{ll}
P_{\nu}(t_{\rm min},t_{\rm j})+P_{\nu}(t_{\rm j},t)\hspace{8mm}{\rm for}\hspace{3mm}t_{\rm j}>t_{\rm min}\\
P_{\nu}(t_{\rm min},t)\hspace{25mm}{\rm for}\hspace{3mm}t_{\rm j}\leq t_{\rm min}, \end{array} \right.
\end{equation}

\noindent
where $t_{\rm min}$ is the injection time of the particles with the largest Lorentz
factors permissible by the model. In the above equation, the first term corresponds
to the integral given by \citet{Kaiser1997b} [their Equation 16].
The second term is given by

\begin{equation}
P_{\nu}(t_{\rm j},t)=\frac{\sigma_{T}c}{6\pi\nu}\frac{r}{r+1}Q_{\rm j}{\cal P}_{\rm hc}
^{(1-\Gamma_{\rm c})/\Gamma_{\rm c}}\int_{t_{0}}^{t}H(t_{\rm i})G(t_{\rm i})dt_{\rm i},
\end{equation}

\noindent
where $t_{0}=t_{\rm min}$ if $t_{\rm j}\leq t_{\rm min}$ and $t_{0}=t_{\rm j}$ if
$t_{\rm j}>t_{\rm min}$,

\[H(t_{\rm i})=n_{0}(t_{\rm i})\frac{\gamma^{3-p}t_{\rm i}^{a_{1}/3(p-2)}}
{[t^{-a_{1}/3}-a_{2}(t,t_{\rm i})\gamma]^{2-p}}\left(\frac{t}{t_{\rm i}}\right)
^{-a_{1}(1/3+\Gamma_{\rm B})},\]

\[G(t_{\rm i})=\]
\[\frac{\int_{0}^{1}F_{\rm JP}(x)x^{-p}(1-x)^{p-2}dx}
{\int_{0}^{1}F_{\rm CI}(x)x^{-(p+1)}[1-(1-x)^{p-2}]dx+\int_{1}^{\infty}F_{\rm JP}(x)x^{-(p+1)}dx},\]

\noindent
where the energy distribution of the injected particles is $n(\gamma)=n_{0}\gamma^{-p}$ and
$p=1+2\alpha_{\rm inj}$.
The functions $F_{\rm JP}=\nu/\nu_{\rm br,JP}/x^{2}$ and $F_{\rm CI}=\nu/\nu_{\rm br,CI}/x^{2}$,
where the frequency breaks in the radio spectrum, are

\[\nu_{\rm br,CI}=\frac{c_{\nu}}{c_{\epsilon}^{2}}\frac{B}{(B^{2}+B_{\rm iC}^{2})^{2}t^{2}},\hspace{3mm}{\rm and}\]

\[\nu_{\rm br,JP}=\frac{c_{\nu}}{c_{\epsilon}^{2}}\frac{B}{(\frac{2}{3}B^{2}+B_{\rm iC}^{2})^{2}(t-t_{\rm j})^{2}},\]

\noindent
$c_{\nu}$ and $c_{\epsilon}$ are physical constants (for details, see \citealt{Pacholczyk1970}),
and $B$ and $B_{\rm iC}$ are the strengths of the magnetic field in the source (cocoon) and the
equivalent field associated with the inverse Compton scattering of the CMB photons, respectively.

\end{document}